\RequirePackage{silence}
\documentclass[fleqn,usenatbib,useAMS]{mnras} 
\usepackage{graphicx}
\let\normalbigoplus \bigoplus
\usepackage{amsmath}
\let\bigoplus \normalbigoplus

\usepackage{amsfonts}
\usepackage{float}
\usepackage{bm}
\usepackage[perpage,symbol*]{footmisc}
\usepackage{ae,aecompl}
\usepackage{array}
\usepackage{soul}
\usepackage{mathtools}
\usepackage{multirow}

\newcommand{\appropto}{\mathrel{\vcenter{
\offinterlineskip\halign{\hfil$##$\cr 
\propto\cr\noalign{\kern2pt}\sim\cr\noalign{\kern-2pt}}}}}

\title{Solar System limits on gravitational dipoles} 

\author[Indranil Banik \& Pavel Kroupa]{Indranil Banik$^{1}$\thanks{Email:
\href{mailto:ibanik@astro.uni-bonn.de}{ibanik@astro.uni-bonn.de} (Indranil Banik)\newline $~~~~~~~~~~~~~~~~~~~~$ \href{mailto:pavel@astro.uni-bonn.de}{pkroupa@uni-bonn.de} (Pavel Kroupa)} and Pavel Kroupa$^{1,2}$\\
$^{1}$Helmholtz-Institut f\"ur Strahlen und Kernphysik (HISKP), University of Bonn, Nussallee 14$-$16, D-53115 Bonn, Germany \\
$^{2}$Charles University, Faculty of Mathematics and Physics, Astronomical Institute, V Hole\v{s}ovi\v{c}k\'ach 2, CZ-180 00 Praha 8, Czech Republic}

\pubyear{2020}
\begin{document}
\label{firstpage}
\pagerange{\pageref{firstpage}--\pageref{lastpage}}

\maketitle

\begin{abstract}


The gravitational dipole theory of Hadjukovic (2010) is based on the hypothesis that antimatter has a negative gravitational mass and thus falls upwards on Earth. Astrophysically, the model is similar to but more fundamental than Modified Newtonian Dynamics (MOND), with the Newtonian gravity $g_{_N}$ towards an isolated point mass boosted by the factor $\nu = 1 + \left( \alpha/x \right) \tanh \left( \sqrt{x}/\alpha \right)$, where $x \equiv g_{_N}/a_{_0}$ and $a_{_0} = 1.2 \times 10^{-10}$ m/s$^2$ is the MOND acceleration constant. We show that $\alpha$ must lie in the range ${0.4-1}$ to acceptably fit galaxy rotation curves. In the Solar System, this interpolating function implies an extra Sunwards acceleration of ${\alpha a_{_0}}$. This would cause Saturn to deviate from Newtonian expectations by ${7000 \left( \alpha/0.4 \right)}$ km over 15 years, starting from known initial position and velocity on a near-circular orbit. We demonstrate that this prediction should not be significantly altered by the postulated dipole haloes of other planets due to the rather small region in which each planet's gravity dominates over that of the Sun. The orbit of Saturn should similarly be little affected by a possible ninth planet in the outer Solar System and by the Galactic gravity causing a non-spherical distribution of gravitational dipoles several kAU from the Sun. Radio tracking of the Cassini spacecraft orbiting Saturn yields a ${5\sigma}$ upper limit of 160 metres on deviations from its conventionally calculated trajectory. These measurements imply a much more stringent upper limit on $\alpha$ than the minimum required for consistency with rotation curve data. Therefore, no value of $\alpha$ can simultaneously match all available constraints, falsifying the gravitational dipole theory in its current form at extremely high significance.


\end{abstract}


\section{Introduction}
\label{Introduction}


One of the great mysteries in contemporary astronomy is the true cause of the very large dynamical discrepancies between the observed rotation curves of galaxies and the predictions of Newtonian gravity applied to their luminous matter distributions \citep[e.g.][]{Babcock_1939, Rubin_Ford_1970, Rogstad_1972}. These acceleration discrepancies are usually attributed to dark matter haloes surrounding each galaxy \citep{Ostriker_Peebles_1973}. This interpretation is challenged by continued null detection of the constituent dark matter particles in sensitive searches, for instance in 11 years of Fermi data on dwarf spheroidal satellites of the Milky Way \citep{Hoof_2020}. Moreover, the galactic acceleration discrepancies follow some remarkable regularities \citep{Famaey_McGaugh_2012} that can be summarised as a unique relation between the acceleration inferred from the rotation curve and that expected from the baryonic distribution \citep{McGaugh_Lelli_2016}.

Such a radial acceleration relation (RAR) was predicted several decades earlier using Modified Newtonian Dynamics \citep[MOND,][]{Milgrom_1983}. In MOND, the supposed dynamical effect of dark matter is instead provided by an acceleration dependence of the gravity law. The gravitational field strength $g$ at distance $r$ from an isolated point mass $M$ transitions from the Newtonian $GM/r^2$ law at short range to
\begin{eqnarray}
	g ~=~ \frac{\sqrt{GMa_{_0}}}{r} ~~~\text{for } ~ r \gg r_{_M} \equiv \sqrt{\frac{GM}{a_{_0}}} \, .
	\label{Deep_MOND_limit}
\end{eqnarray}
MOND introduces $a_{_0}$ as a fundamental acceleration scale of nature below which the deviation from Newtonian dynamics becomes significant. For a point mass, this corresponds to distances beyond its MOND radius $r_{_M}$.

Empirically, $a_{_0} = 1.2 \times {10}^{-10}$ m/s$^2$ to match galaxy rotation curves \citep{Begeman_1991, McGaugh_2011}. Remarkably, this is the same order of magnitude as the acceleration at which the classical energy density of a gravitational field \citep[equation 9 of][]{Peters_1981} becomes comparable to the dark energy density $u_{_\Lambda} \equiv \rho_{_\Lambda} c^2$ that conventionally explains the accelerating expansion of the Universe \citep{Efstathiou_1990, Ostriker_Steinhardt_1995, Riess_1998}.
\begin{eqnarray}
	\frac{g^2}{8\mathrm{\pi}G} ~<~ u_{_\Lambda} ~~\Leftrightarrow~~ g ~\la~ 2\mathrm{\pi}a_{_0} \, .
	\label{MOND_quantum_link}
\end{eqnarray}
MOND could thus be a result of poorly understood quantum gravity effects that also underlie dark energy \citep[e.g.][]{Milgrom_1999, Pazy_2013, Verlinde_2016, Smolin_2017}.

Regardless of the microphysical explanation, MOND can accurately match the rotation curves of a wide variety of spiral and elliptical galaxies across a vast range in mass, surface brightness and gas fraction using only their luminous matter \citep[fig. 5 of][]{Lelli_2017}. This only became apparent long after the MOND field equation was first published \citep{Bekenstein_Milgrom_1984}, making these achievements successful a priori predictions. It is difficult to explain the success of these predictions in a conventional gravity context, even with the observational facts in hand \citep{Desmond_2016, Desmond_2017, Ghari_2019}. In particular, the latter work showed that it is still very difficult to keep a tight RAR despite a diversity of rotation curve shapes at fixed peak velocity \citep{Oman_2015}. 

Despite its successes, the empirical MOND approach lacks a firm theoretical foundation. The gravitational dipole (GD) model seeks to explain the MOND phenomenology using more fundamental concepts \citep{Hajdukovic_2010}. In this model, antimatter particles have a negative gravitational mass with the same magnitude as the corresponding particle. The gravitational properties of antimatter are under investigation at CERN thanks to the experiments known as GBAR \citep{Perez_2015}, AEgIS \citep{Brusa_2017} and ALPHA-g \citep{Bertsche_2018}. Experimental uncertainties are currently $\approx 100 \times$ too large to test the hypothesis that antimatter falls upwards in the terrestrial gravitational field \citep{Charman_2013}. Decisive results should be available by the end of 2021.

If antimatter anti-gravitates in the sense described, matter-antimatter pairs resulting from quantum vacuum (QV) fluctuations would be virtual GDs. An important consequence is that the QV contains no net gravitational mass, leading to a predicted cosmological constant of zero. In the GD model, the dynamical discrepancies in galaxies are an effect of gravitational polarization of the GDs. The divergence of this polarized field of GDs leads to extra effective mass, similarly to polarization in electrostatics (Equation \ref{rho_qv}).

In this contribution, we focus on what the GD model implies for the Solar System. The accurate data within it place tight constraints on modified gravity theories \citep{Hees_2016}. In the context of MOND as modified gravity, this is mainly due to the Galactic external field effect \citep{Milgrom_1986} causing the Solar `phantom' dark matter halo to be non-spherical at large distances. Thus, even without any phantom dark matter within the inner Solar System, deviations from Keplerian motion are generically expected due to a non-zero tidal stress \citep{Blanchet_2011}.

The GD model originally predicted a fixed extra Sunwards acceleration of $\approx 7 a_{_0}$ in the Solar System \citep[equation 12 of][]{Hajdukovic_2010}. This was in line with the `Pioneer anomaly', a then-unexplained Sunwards acceleration of the Pioneer spacecraft \citep{Turyshev_2010}. However, this anomaly is likely caused by thermal forces \citep{Turyshev_2012}. The expected anomalous acceleration was later revised down to $a_{_0}$ \citep{Hajdukovic_2012} and then to the range ${\left( 0.1-1 \right)} a_{_0}$ \citep[equation 6 of][]{Hajdukovic_2013}. A theoretically motivated interpolating function was then provided between the `saturated' (high-acceleration) region near a massive body and the low-acceleration region further away, where the extra gravity of the GDs is expected to dominate \citep[section 2 of][]{Hajdukovic_2014}. We make use of equation 2 in that work, which was also restated in equation 24 of \citet{Hajdukovic_final}. These equations are analogous to our Equation \ref{Force_law_Hadjukovic}. Importantly for our work, the latest publication advocating the GD model predicts that Solar System planets should experience an extra Sunwards acceleration of $\left(0.4 - 0.5 \right) a_{_0}$, with the most likely value being $5 a_{_0}/12$ \citep[section 6.1 of][]{Hajdukovic_2020}.

The effects of radiation pressure can be minimized by accurately tracking the motion of a gas giant planet, as done by the Cassini mission \citep{Matson_1992}. Using radio tracking data from orbit around Saturn, \citet{Hees_2014} constrained the tidal stress on the Solar System to be ${\left( 3 \pm 3 \right)} \times 10^{-27}/s^2$ in the direction towards the Galactic Centre (see their equation 6). At the 9.58 AU distance of Saturn, this corresponds to an acceleration of ${\left( 4.3 \pm 4.3 \right)} \times 10^{-15}$ m/s$^2$. While this constraint is not directly applicable to an additional Sunwards force, it does suggest that any such anomalous acceleration should be $\la 10^{-4} a_{_0}$. This casts serious doubt on the GD model \citep{Iorio_2019}. We investigate the model in more detail to see if there is any way around this apparent falsification.

After introducing the GD model and its context (Section \ref{Introduction}), we explore what a GD force law consistent with galaxy rotation curves (Section \ref{Constraints}) implies for the motion of a planet in the Solar System (Section \ref{Solar_System}). We find that the GD model predicts very large deviations  (Sections \ref{Analytic_estimates}-\ref{Epicyclic_approximation}) that greatly exceed observational upper limits (Section \ref{Observations}), as previously found by \citet{Iorio_2019}. We then explore whether this conclusion can be weakened once we take into account the gravity of other planets and their dipole haloes (Section \ref{Planetary_perturbations}), the possible presence of a ninth planet in the outer Solar System (Section \ref{Planet_9}), and the Galactic gravity (Section \ref{Galactic_gravity}). Our conclusions are given in Section \ref{Conclusions}.

\section{Galaxy rotation curve constraint}
\label{Constraints}

In this section, we consider galaxy rotation curve constraints on the isolated point mass force law in the GD model, which can be expressed as \citep[equation 24 of][]{Hajdukovic_final}:
\begin{eqnarray}
	g ~=~ g_{_N} + \, \alpha a_{_0} \tanh \left( \frac{r_{_M}}{\alpha r}\right) \, .
	\label{Force_law_Hadjukovic}
\end{eqnarray}
This reduces to Newtonian dynamics when the Newtonian gravity $g_{_N} \equiv GM/r^2 \gg a_{_0}$. It also reproduces the correct asymptotic behaviour for $g_{_N} \ll a_{_0}$ (Equation \ref{Deep_MOND_limit}).

Equation \ref{Force_law_Hadjukovic} can be written in the standard MOND terminology with interpolating function
\begin{eqnarray}
	\nu_{_{GD}} ~=~ 1 \, + \, \frac{\alpha}{x} \tanh \left( \frac{\sqrt{x}}{\alpha}\right) ,
	\label{Interpolating_function_Hadjukovic}
\end{eqnarray}
where $x \equiv g_{_N}/a_{_0}$ and the actual gravity $g ~\equiv~ \nu g_{_N}$, with $\nu$ being the factor by which a model enhances the strength of gravity towards an isolated point mass. The subscript $_{GD}$ indicates that Equation \ref{Interpolating_function_Hadjukovic} is motivated by the GD model.

\begin{figure}
	\centering
	\includegraphics[width = 8.5cm] {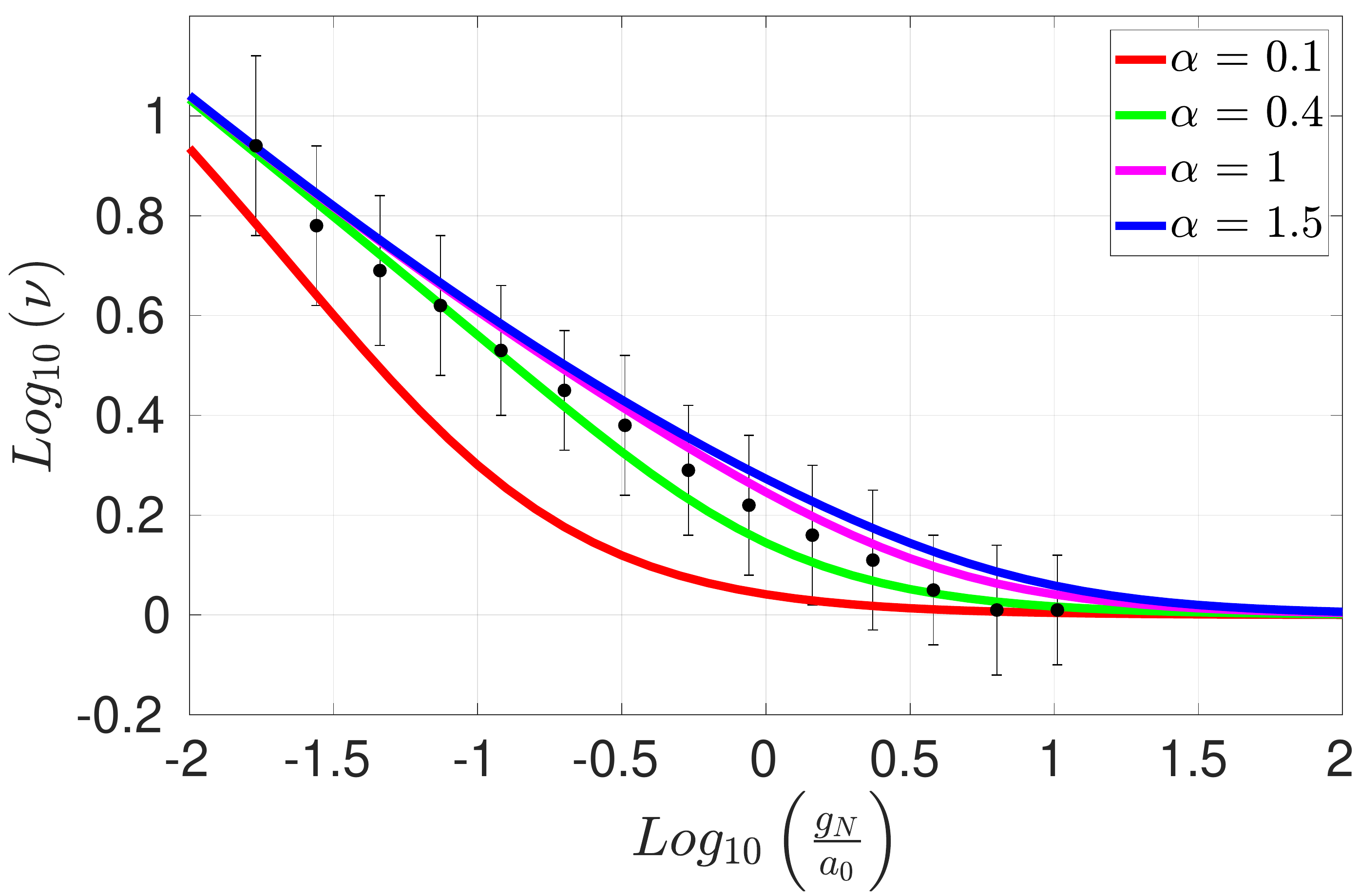}
	\includegraphics[width = 8.5cm] {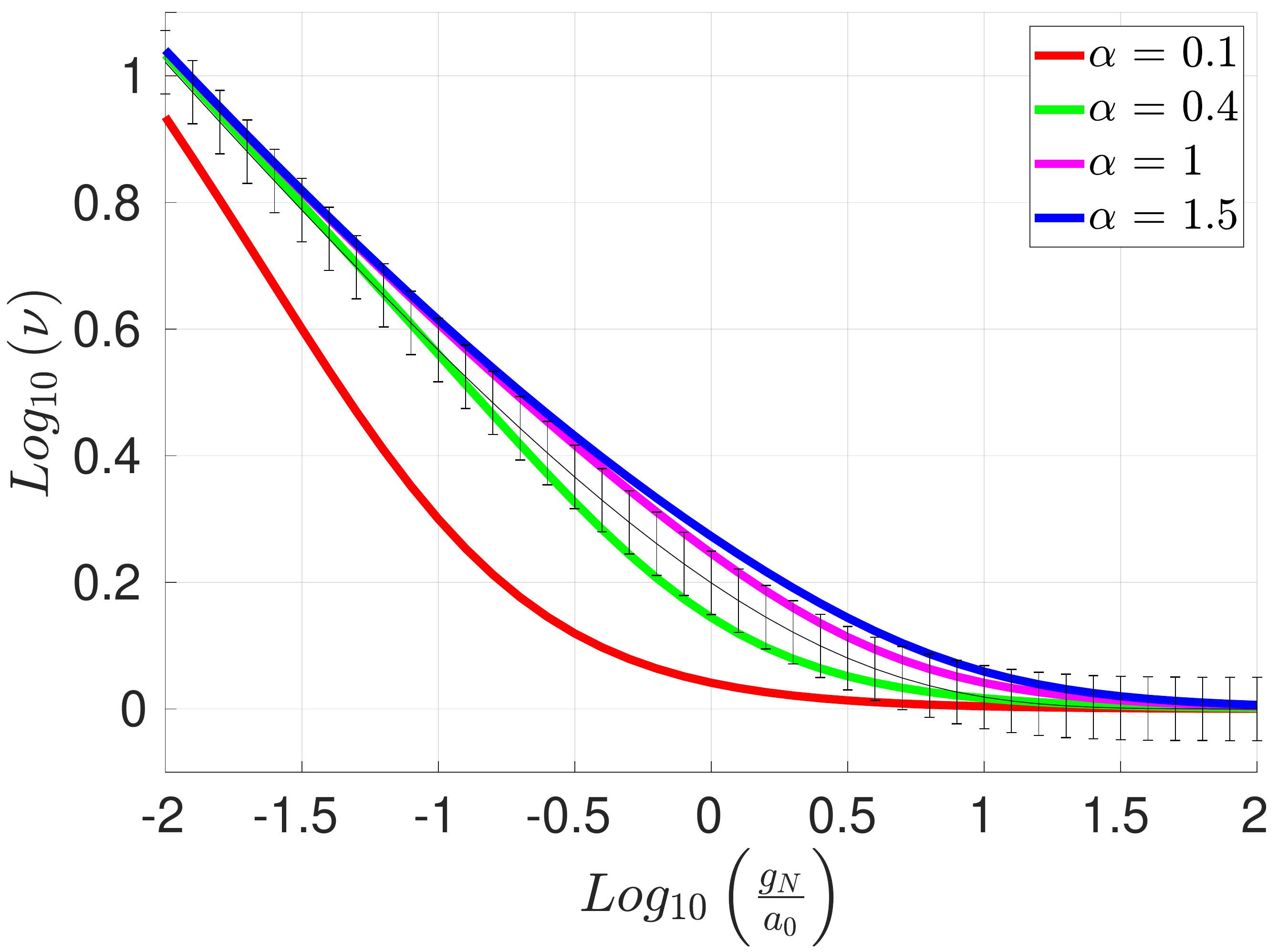}
	\caption{Dipole dark matter-based interpolating functions (Equation \ref{Interpolating_function_Hadjukovic}) for different values of $\alpha$ (curves with higher $\alpha$ have higher $\nu$). We also show \emph{top}: the binned RAR data from \citet{Lelli_2017} with error bars showing (conservatively) the dispersion in $\log_{10} g$ for each bin, and \emph{bottom}: the empirical fit to their data (Equation \ref{nu_RAR}) with 0.05 dex error bars, based on fits to galaxy rotation curves \citep{Li_2018}. It is clear that $\alpha$ should lie in the range $\left( 0.4-1 \right)$, with $\alpha < 0.1$ strongly excluded.}
	\label{Interpolating_functions}
\end{figure}

The $\nu$ function is constrained by the baryonic mass distributions and rotation curves of disc galaxies, with the former determining $g_{_N}$ while the latter measures the true radial gravity $g$. The empirical relation between $g$ and $g_{_N}$ (the RAR) can be fit very well by \citep[equation 11 of][]{Lelli_2017}:
\begin{eqnarray}
	\nu_{_{RAR}} &=& \frac{1}{1 - \mathrm{e}^{-\sqrt{x}}} \, .
	\label{nu_RAR}
\end{eqnarray}
This RAR-inspired interpolating function is rather similar to the `simple' form $\nu_s = \frac{1}{2} + \sqrt{\frac{1}{4} + \frac{1}{x}}$ \citep{Famaey_Binney_2005} for $x \la 10$ \citep[e.g. fig. 1 of][]{Chae_2019}. The main difference is that the exponential term causes much faster convergence to Newtonian dynamics at large $x$, minimizing the impact on the Solar System.

The RAR has an intrinsic scatter $< 0.05$ dex once observational uncertainties are taken into account \citep{Li_2018}. We take this as the uncertainty in the mean relation i.e. in the value of $\nu$. Figure \ref{Interpolating_functions} shows the binned RAR data \citep{Lelli_2017} and their fit (Equation \ref{nu_RAR}) with 0.05 dex error bars for $g_{_N} = \left( 0.01  - 100 \right) a_{_0}$, the approximate range of the dataset. Any viable modification to gravity should have an interpolating function which passes through this region. We also show $\nu_{_{GD}}$ for different values of the parameter $\alpha$ (Equation \ref{Interpolating_function_Hadjukovic}). To match constraints from galaxy rotation curves, ${\alpha = 0.4-1}$ since $a_{_0} = 1.2 \times {10}^{-10}$ m/s$^2$ is fixed by the relation between galaxy baryonic masses and flatline rotation curve levels \citep{Begeman_1991}.

\section{Solar System constraint}
\label{Solar_System}

Equation \ref{Deep_MOND_limit} indicates that the MOND radius of the Sun is $r_{_M} = 7000$ astronomical units (7 kAU), implying that the whole Solar System lies well inside $r_{_M}$ and is not much affected by MOND. However, the vastly superior data quality within the Solar System makes it important to explore what the GD model predicts for $g_{_N} \gg a_{_0}$. We focus on spacecraft tracking data at heliocentric distances $\la 10$ AU. Planets in this region should feel an extra Sunwards acceleration of $\alpha a_{_0}$ (Equation \ref{Force_law_Hadjukovic}). This aspect of the dipole model was also discussed in section 4.1 of \citet{Hajdukovic_2020} and in its abstract, where it was claimed that such a small effect (estimated at $6 \times {10}^{-11}$ m/s$^2$) cannot be detected in the Solar System. We will see that this is incorrect.

\subsection{Order of magnitude estimate}
\label{Analytic_estimates}

After some observing duration $t$, the anomalous acceleration of $\alpha a_{_0}$ implies a position drift of $\approx \alpha a_{_0}t^2/2$ starting from the same initial position and velocity. For $\alpha = 0.4$, the expected deviation from Keplerian motion is 600~km after 5 years for a planet like Saturn with a much longer orbital period.

\citet{Iorio_2019} first pointed out that this casts serious doubt on the GD model since radio tracking data for the Cassini spacecraft has an accuracy of 32 m, but we have not detected such large deviations from its conventionally calculated trajectory \citep[table 11 of][]{Viswanathan_2017}.

\subsection{Epicyclic approximation}
\label{Epicyclic_approximation}

We now consider how a planet's trajectory would be affected by a small non-Newtonian force. Our derivation will be similar to that in \citet{Banik_DMwake}.

Consider a planet on a near-circular orbit around the Sun at heliocentric radius $r_p$. Its Keplerian orbital frequency $\Omega = \sqrt{GM_\odot/{r_p}^3}$. Small radial perturbations $r$ can be understood using the epicyclic approximation $\ddot{r} = -\Omega^2 r$.\footnote{Epicyclic and orbital frequencies are the same only for an inverse square central force.} Since this is linear in $r$, we may subtract whatever epicyclic oscillations the planet is undergoing and concentrate on the evolution of any remaining anomaly $\delta r$. This anomaly is sensitive to any additional non-Keplerian forces acting on the planet. Here, we focus on the extra Sunwards acceleration of $\alpha a_{_0}$ in the GD model.
\begin{eqnarray}
	\ddot{\delta r} ~=~ -\Omega^2 \delta r - \alpha a_{_0} \, .
	\label{Perturbation_r_governing_equation}
\end{eqnarray}
This should be solved subject to the condition $\delta r = \dot{\delta r} = 0$ at some initial time $t = 0$ when precise observations of the planet start. There is no way to know if its position and velocity have been altered by non-Keplerian forces at earlier times. Such forces are relevant only in so far as they cause the planet to deviate from the Keplerian trajectory corresponding to its osculating orbital elements at $t = 0$. Thus, the relevant solution to Equation \ref{Perturbation_r_governing_equation} is
\begin{eqnarray}
	\delta r &=& x_0 \left( \cos \phi - 1 \right), \, \text{where} ~\phi ~\equiv~ \Omega t ~ \text{and} \\
	x_0 &=& \frac{\alpha a_{_0}}{\Omega^2} ~=~ \frac{\alpha a_{_0} {r_p}^3}{G M_\odot} \, .
	\label{x_0}
\end{eqnarray}
For Saturn orbiting at $r_{_{Sat}} = 9.58$~AU, $x_0 = 1065$~km.

The GD model yields an extra Sunwards force, implying angular momentum is still conserved. Thus, the angular velocity $\propto \left( r_p + \delta r \right)^{-2}$, leading to a perturbation $\delta p$ parallel to the circular orbital velocity. This is governed by
\begin{eqnarray}
	\dot{\delta p} ~=~ -2 \Omega \delta r \, .
	\label{Perturbation_y_governing_equation}
\end{eqnarray}
For reasons just explained, we also require $\delta p = \dot{\delta p} = 0$ at $t = 0$, though the latter is guaranteed if $\delta r = 0$ then. Thus, the relevant solution to Equation \ref{Perturbation_y_governing_equation} is
\begin{eqnarray}
	\delta p ~=~ 2 x_0 \left(\phi - \sin \phi \right) \, .
\end{eqnarray}

\begin{figure}
	\centering
		\includegraphics[width = 8.5cm] {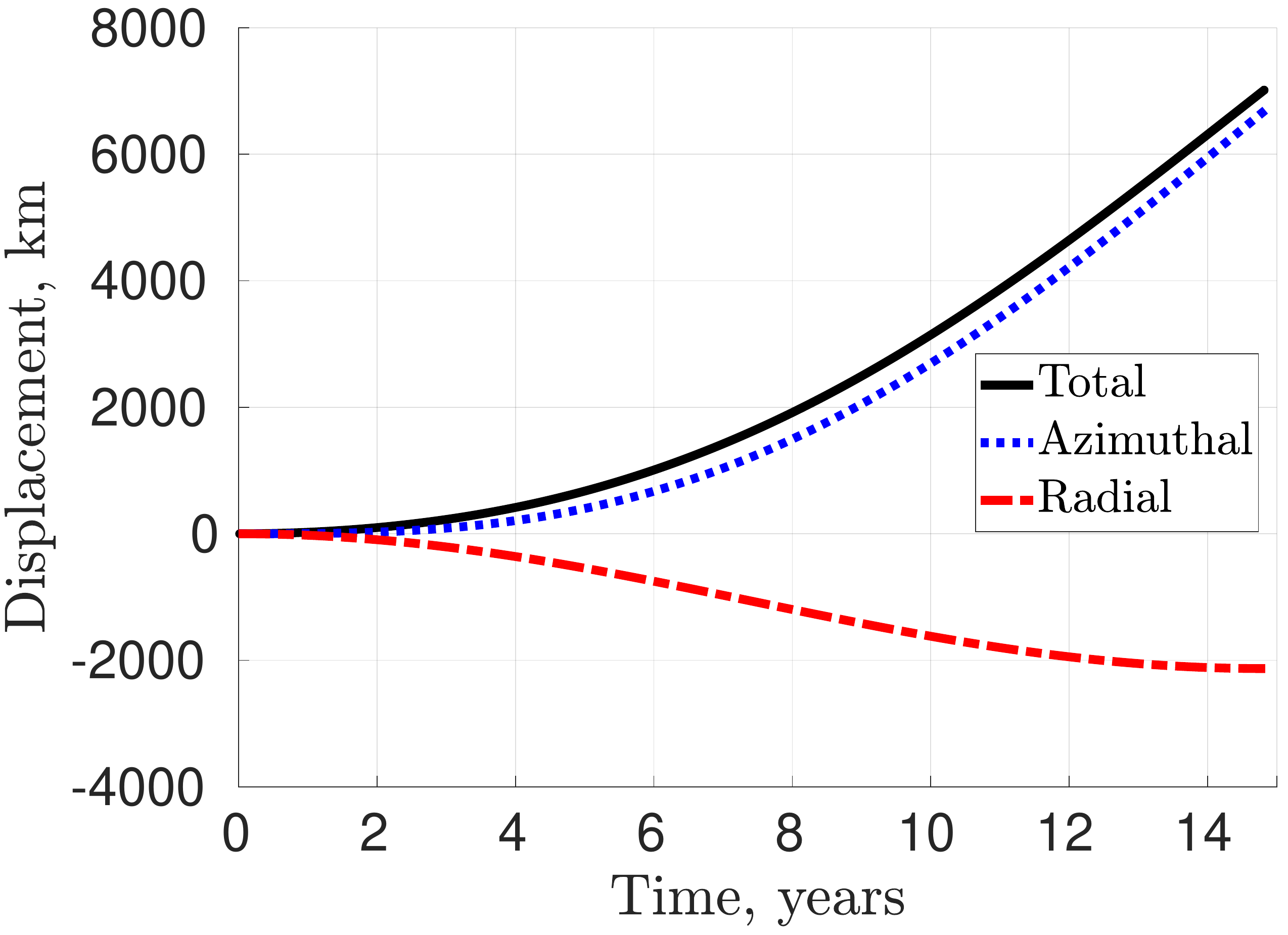}
		\caption{The difference between Newtonian and GD predictions (former subtracted) for the position of Saturn if $\alpha = 0.4$, the minimum needed for consistency with galaxy rotation curves (Figure \ref{Interpolating_functions}). The displacements are $\propto \alpha$. Azimuthal displacements are positive in the direction Saturn rotates around the Sun. The much smaller orbit of Earth is much less affected by GDs (Equation \ref{x_0}). Thus, the difference between the models in the Earth-Saturn range is approximately that in Saturn's heliocentric distance.}
	\label{Displacement}
\end{figure}

The orbital period of Saturn is ${\approx 30}$ years and the Cassini spacecraft orbited it for almost half this time. The GD-induced displacement of Saturn when $\phi = \mathrm{\pi}$ is
\begin{eqnarray}
	\sqrt{\delta r^2 + \delta p^2} ~=~ 2 x_0 \sqrt{1 + \mathrm{\pi}^2} ~=~ 7025 \, \text{km}.
	\label{Saturn_displacment}
\end{eqnarray}
The displacement at earlier times is illustrated in Figure \ref{Displacement}.

\subsection{Comparison with observations}
\label{Observations}

The latest planetary ephemerides for Saturn are accurate to 32 m \citep[table 11 of][]{Viswanathan_2017} thanks to radio tracking data from the Cassini spacecraft, which orbited Saturn for 13 years \citep{Matson_1992}. The null detection of deviations from its conventionally calculated trajectory implies $\alpha \la 10^{-5}$. Since $\alpha \la 0.4$ violates galaxy rotation curve data (Figure \ref{Interpolating_functions}), the GD model must be discarded in its present form. This conclusion was previously reached by \citet{Iorio_2019} based on the rate of perihelion precession, but frequent radio tracking of Cassini provides much more information than just this time-averaged quantity.

Our results completely rule out MOND interpolating functions of the form in Equation \ref{Interpolating_function_Hadjukovic}. This form is fundamental to the whole GD approach \citep[section 2 of][]{Hajdukovic_2014}. In particular, the $\tanh$ function is derived directly from the partition function of the dipoles \citep[section 4.2 of][]{Hajdukovic_2020}. In the remainder of this contribution, we explore the GD model in more detail to check if it contains features that might weaken this falsification.



\subsection{Effects of other planets and their dipole haloes}
\label{Planetary_perturbations}

In the Solar System, the GDs would be fully saturated as $g_{_N} \gg a_{_0}$. This approximation $-$ also made in section 5.1 of \citet{Hajdukovic_2020} $-$ implies that the polarization density $\bm{P}$ has its maximum strength of $P_{max}$ and is aligned with the gravitational field $\bm{g}$. Thus, the QV effective density
\begin{eqnarray}
	\label{rho_qv}
	\rho_{qv} ~=~ - \nabla \cdot \bm{P}, ~ \text{where} ~ \bm{P} ~=~ P_{max} \, \widehat{\bm{g}} ~ \left( g \gg a_{_0} \right).
\end{eqnarray}
We use the convention that $v \equiv \left| \bm{v} \right|$ for any vector $\bm{v}$ and $\widehat{\bm{v}}$ is the unit vector in the same direction. In the Solar System, GDs have only a small effect on $\bm{g}$, so $\widehat{\bm{g}} \approx \widehat{\bm{g}}_{_N}$. To get an extra Sunwards acceleration of $\alpha a_{_0}$, we must have that
\begin{eqnarray}
	P_{max} ~=~ \frac{\alpha a_{_0}}{4 \mathrm{\pi} G} \, .
	\label{P_max}
\end{eqnarray}

We now find the total mass of the GD halo associated with a planet of mass $M_p$ embedded in an external gravitational field $\bm{g}_{_{ext}}$ from the Sun. The Solar gravity is assumed to be uniform over the whole planetary halo, an assumption we will justify later.

\begin{figure}
	\centering
		\includegraphics[width = 8.5cm] {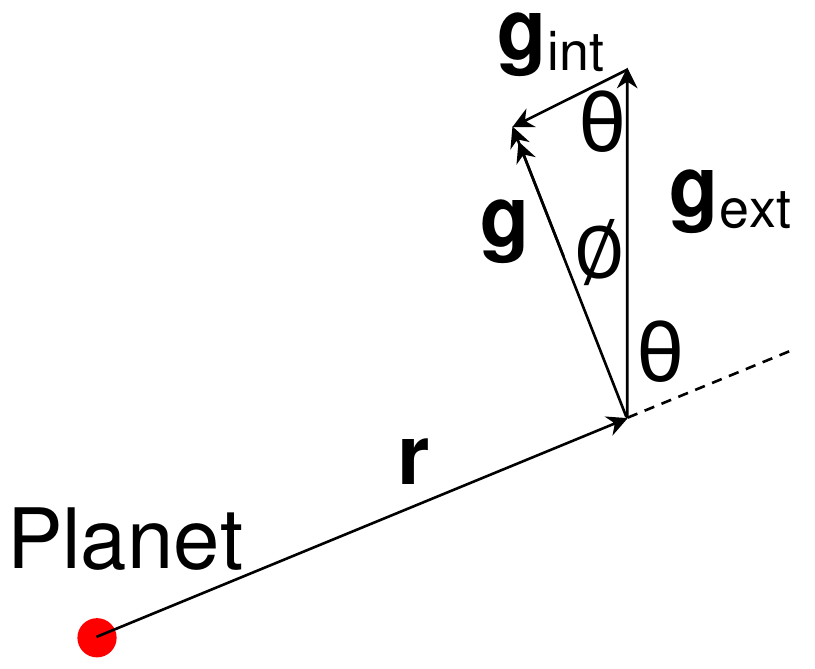}
		\caption{The total gravitational field $\bm{g}$ at planet-centric position $\bm{r}$ is the vector sum of the external Solar gravity $\bm{g}_{_{ext}}$ and the planetary gravity $\bm{g}_{_{int}}$, which deflects the direction of $\bm{g}$ by a small angle $\phi$ in the region where the Solar gravity dominates ($g_{_{int}} \ll g_{_{ext}}$). $\theta$ is the angle between $\bm{g}_{_{ext}}$ and $\bm{r}$.}
	\label{Dipole_diagram}
\end{figure}

We use spherical polar co-ordinates $\left( r, \theta \right)$ centred on the planet with axis parallel to $\widehat{\bm{g}}_{_{ext}}$ i.e. towards the Sun. Applying the divergence theorem to a spherical region of radius $r$ centred on the planet, we get that the volume integral
\begin{eqnarray}
	\int \rho_{qv} \, dV ~=~ -P_{max} \int_0^\mathrm{\pi} \left( \widehat{\bm{g}} \cdot \widehat{\bm{r}} \right) 2 \mathrm{\pi} r^2 \sin \theta \, d\theta \, .
	\label{Integrand}
\end{eqnarray}
We can simplify this by taking $r$ to be sufficiently large such that the Solar gravity $\bm{g}_{_{ext}}$ dominates over the planetary gravity $\bm{g}_{_{int}} = -GM_p\widehat{\bm{r}}/r^2$. Without the planetary gravity, $\widehat{\bm{g}} = \widehat{\bm{g}}_{_{ext}}$ and so $\widehat{\bm{g}} \cdot \widehat{\bm{r}} = \cos \theta$, implying $\int \rho_{qv} \, dV = 0$. Thus, we need the first-order perturbation to $\widehat{\bm{g}}$ caused by $\bm{g}_{_{int}}$. The situation is illustrated in Figure \ref{Dipole_diagram}.

As $g_{int} \ll g_{_{ext}}$, the planetary gravity rotates $\widehat{\bm{g}}$ towards the planet by a small angle
\begin{eqnarray}
	\phi ~\approx~ \frac{g_{int} \sin \theta}{g_{_{ext}}} ~=~ \frac{GM_p \sin \theta}{r^2 g_{_{ext}}} \, .
\end{eqnarray}
Since $\phi \ll 1$, we get that
\begin{eqnarray}
	\widehat{\bm{g}} \cdot \widehat{\bm{r}} ~\equiv~ \cos \left( \theta + \phi \right) ~\approx~ \cos \theta - \frac{GM_p \sin^2 \theta}{r^2 g_{_{ext}}} \, .
\end{eqnarray}
Substituting this into Equation \ref{Integrand} and using $\int_0^\mathrm{\pi} \sin^3 \theta \, d\theta = 4/3$, we obtain a planetary QV halo mass of
\begin{eqnarray}
	M_{halo, p} ~=~ \frac{8 \mathrm{\pi} G M_p P_{max} }{3 g_{_{ext}}} \, .
	\label{Halo_mass}
\end{eqnarray}
This result is independent of $r$, implying that the GD halo has a finite mass. This is similar to how the MOND phantom dark matter halo is truncated in the presence of an external field \citep[e.g.][]{Wu_2008}. The $r$-independence of our result shows that the effective mass does not lie in the region where $g_{int} \ll g_{ext}$, so it must lie closer to the planet in the region where its gravity dominates. This means that the planetary GD halo has an approximate size of $r_h$, defined by the distance where $g_{int} = g_{ext}$. A planet at heliocentric distance $r_p$ has
\begin{eqnarray}
	r_h ~=~ r_p \sqrt{\frac{M_p}{M_\odot}} \, .
\end{eqnarray}
For Jupiter, $r_h = 0.16$ AU. As other Solar System planets have even less mass, their GD haloes are concentrated within a very small fraction of their orbital radius. This justifies our assumption that the Solar gravity can be treated as uniform over the planetary halo.

Substituting in Equation \ref{P_max} and the fact that $g_{ext} = GM_\odot/{r_p}^2$, Equation \ref{Halo_mass} can be written as
\begin{eqnarray}
	M_{halo, p} ~=~ \frac{2 \alpha a_{_0} M_p {r_p}^2}{3 G M_\odot} ~=~ \frac{2}{3} \alpha M_p \left( \frac{r_p}{r_{_M}} \right)^2 .
	\label{Planet_halo_mass}
\end{eqnarray}
The Solar GD halo interior to the planetary orbit has an effective mass of
\begin{eqnarray}
	M_{halo, \odot} \left( < r_p \right) ~=~ \frac{\alpha a_{_0} {r_p}^2}{G} ~=~ \alpha M_\odot \left( \frac{r_p}{r_{_M}} \right)^2 .
	\label{Solar_halo_mass}
\end{eqnarray}
As the Solar dipole halo is spherically symmetric, a planet is not affected by $\rho_{qv}$ outside its orbit (though see Section \ref{Galactic_gravity}). We can combine Equations \ref{Planet_halo_mass} and \ref{Solar_halo_mass} to find the ratio between the Jovian effective halo mass and that of the Solar halo interior to the orbit of Saturn.
\begin{eqnarray}
	\frac{M_{halo, Jup}}{M_{halo, \odot} \left( < r_{_{Sat}} \right)} ~=~ \frac{2M_{Jup}}{3M_\odot} \left( \frac{r_{_{Jup}}}{r_{_{Sat}}} \right)^2 =~ 1.9 \times 10^{-4} \, .
\end{eqnarray}
Clearly, the Solar dipole halo is several thousand times more important to Saturn than the dipole haloes of other planets.

This result can be understood by approximating that the planetary gravity is completely dominant inside the region $r < r_h$ and completely negligible outside, such that $\rho_{qv} \neq 0$ only if $r < r_h$. In this case, we would get that
\begin{eqnarray}
	M_{halo, p} \approx 4 \mathrm{\pi} {r_h}^2 P_{max} = \frac{\alpha a_{_0} M_p {r_p}^2}{G M_\odot} = \alpha M_p \left( \frac{r_p}{r_{_M}} \right)^2.
	\label{M_halo_p}
\end{eqnarray}
This quick estimate agrees rather well with our result in Equation \ref{Planet_halo_mass}. The halo mass $\propto {r_h}^2$ because the total amount of GDs scales with volume but it is their divergence which is relevant, leading to an additional $1/r_h$ geometric factor. This is why the GD theory predicts a fixed extra acceleration towards a point mass in the saturated regime (Equation \ref{Force_law_Hadjukovic}). As a result, Jupiter's dipole halo has an effective mass $\approx \left( 0.16/9.6 \right)^2 \times$ smaller than that in the Sun's dipole halo interior to the 9.6 AU radius of Saturn's orbit.

Therefore, the anomalous Sunwards acceleration of Saturn is almost entirely determined by the Solar GD halo. Neglecting the haloes of other planets leads to a fractional uncertainty of order $10^{-4}$, while neglecting its own halo is justified by the fact that a planet should not be accelerated by its own halo. Moreover, increasing the effective mass of Jupiter (or any other planet) yields an extra force towards the planet rather than towards the Sun. Hence, an extra Sunwards acceleration of $\alpha a_{_0}$ cannot be cancelled out by arbitrarily adjusting the effective masses of other planets.

The only possible exception is if we adjust the masses of the terrestrial planets, as one could argue that these are in a similar direction to the Sun as perceived from Saturn. Equation \ref{Solar_halo_mass} implies that we would need to reduce the mass of e.g. Venus by $\Delta M = 0.25 M_\bigoplus$, almost $5\times$ the mass of Mercury. We would need a different $\Delta M$ in order to correctly `tune' the orbit of a different gas giant since $M_{halo, \odot} \left( < r_p \right) \propto {r_p}^2$. In any case, such large values of $\Delta M$ greatly exceed observational uncertainties because all known Solar System planets have rather well-known masses thanks to data from close spacecraft flybys. This also applies to the Sun $-$ the Solar gravitational parameter $GM_\odot$ has been measured to an accuracy of $2.5 \times 10^{-4} G M_\bigoplus$ \citep{Pitjeva_2015}.

\citet{Hajdukovic_2019} claim that including the gravity of dipoles which ``do not belong to any individual halo'' might alleviate the above issues. Since unpolarized (randomly aligned) GDs have no divergence and thus contribute nothing to $\rho_{qv}$ (Equation \ref{rho_qv}), this might refer to regions where the gravity from a planet is comparable to that from the Sun. It is precisely these regions which we have just analysed in detail. There is another situation where it is difficult to identify the dominant object $-$ at heliocentric distances of several kAU, the Galactic gravity becomes important. In Section \ref{Galactic_gravity}, we consider the possible impact of GDs in this region on the orbits of Solar System planets.



\subsection{Additional planets in the outer Solar System}
\label{Planet_9}

An undiscovered ninth planet (P9) several hundred AU from the Sun \citep{Batygin_2016} would raise tides on the known Solar System, affecting planetary orbits. The tidal effect of P9 on the Sun-Saturn relative acceleration is
\begin{eqnarray}
	g_{tide} ~\approx~ \frac{GM_{_9}r_{_{Sat}}}{{r_{_9}}^3} \, ,
\end{eqnarray}
where $r_9$ is the present heliocentric distance of P9. For $g_{tide}$ to be strong enough to correctly tune the orbit of Saturn, it must be comparable to $\alpha a_{_0}$, implying that
\begin{eqnarray}
	M_{_9} ~\approx~ \frac{\alpha a_{_0} {r_{_9}}^3}{Gr_{_{Sat}}} ~=~ 281 M_\bigoplus \left( \frac{r_{_9}}{100 \, \text{AU}}\right)^3 \, .
	\label{Required_M9}
\end{eqnarray}
However, the proposed P9 has a mass of only ${M_9 \approx 10 \, M_\bigoplus}$ and is more distant \citep[${r_9 \ga 200}$ AU,][]{Batygin_2016}. In fact, the latest analysis indicates that P9 must be ${\ga 370}$ AU away if ${M_9 = 10 \, M_\bigoplus}$ \citep[section 2.6 of][]{Batygin_2019}. Such a distant planet should create only a very small tidal effect on the known Solar System.

In the GD model, part of the effective mass of P9 would come from its dipole halo. However, Equation \ref{M_halo_p} shows that
\begin{eqnarray}
	\frac{M_{halo, p}}{M_p} \, = \, \frac{\alpha a_{_0} {r_p}^2}{G M_\odot} \, = \, 5.4 \times 10^{-5} \left( \frac{r_p}{100 \, \text{AU}} \right)^2 \left( \frac{\alpha}{0.4} \right).
\end{eqnarray} 
Moreover, the $M_9$ estimate in \citet{Batygin_2016} should already include its dipole halo since it is based on how P9 perturbs the orbits of distant Kuiper Belt Objects.

Even if tides from P9 could affect the Sun-Saturn relative acceleration by $\alpha a_{_0}$, P9 would not always pull Saturn directly away from the Sun over a 15 year period. Cassini tracking data would easily reveal any lateral force of order $\alpha a_{_0}$ maintained for a few years (Section \ref{Analytic_estimates}). Moreover, extreme fine-tuning would be required for P9 to have the right tidal effect on the orbit of Saturn (Equation \ref{Required_M9}).

The effect would be different on other planets as it would scale with the size of their orbit. However, the GD model predicts that they too should feel an extra Sunwards acceleration of $\alpha a_{_0}$ due to the Solar dipole halo interior to their orbit (Equation \ref{Force_law_Hadjukovic}). This means that different P9 masses are required to counter this effect for different planets. Consequently, the possible existence of P9 cannot reconcile the GD model with observations.

\subsection{Effect of the Galactic gravity}
\label{Galactic_gravity}

The Solar dipole halo exterior to the orbit of Saturn does not affect its dynamics because the halo is spherically symmetric. This symmetry eventually breaks down to mere axisymmetry at distances $\ga 5$~kAU due to the Galactic gravitational field. As a result, $\rho_{qv}$ in this region can raise tides on the Solar System. Since the Galactic gravity on the Sun is of order $a_{_0}$ \citep[e.g. section 3.6 of][]{Banik_2018_Centauri}, we expect order unity deviations from Newtonian dynamics at the Sun's MOND radius of 7~kAU (Equation \ref{Deep_MOND_limit}). In other words, we expect tides from the non-spherical $\rho_{qv}$ distribution to have a strength of order $a_{_0}$ at this distance. Given that Saturn is $<10$ AU from the Sun, the Sun-Saturn relative acceleration should only be affected by ${\la 10^{-3} a_{_0}}$.

Even if these tides were much stronger, tides raised by such distant $\rho_{qv}$ cannot always cause a Sun-Saturn repulsion while Saturn completes half an orbit. This is because the tidal field due to $\rho_{qv}$ outside the Solar System must be divergence free within the Solar System. Repulsive tides can be maintained around half a Saturnian orbit only if the $\rho_{qv}$ distribution is axisymmetric and the symmetry axis aligns with the orbital pole of Saturn. However, there is a $95.5^\circ$ angle between the North Ecliptic Pole and the direction towards the Galactic Centre. Therefore, tides raised by the non-spherical $\rho_{qv}$ distribution several kAU from the Sun cannot cancel a constant extra Sunwards acceleration of order $a_{_0}$. Moreover, tidal accelerations scale with the size of the planetary orbit while the extra Sunwards acceleration in the GD model is the same for all planets, making it infeasible for tides to consistently cancel out this effect.

\section{Conclusions}
\label{Conclusions}

The GD theory \citep{Hajdukovic_2010} is similar to MOND with an interpolating function given by Equation \ref{Interpolating_function_Hadjukovic}. It seeks to explain the MOND phenomenology by postulating that antimatter has a negative gravitational mass and so falls upwards in the terrestrial gravitational field. This should be directly testable in the next few years.

The model also predicts that Solar System planets experience an extra Sunwards acceleration of ${\alpha a_{_0}}$, where the MOND parameter $a_{_0} = 1.2 \times {10}^{-10}$ m/s$^2$ in order to match rotation curve constraints. These require $\alpha \ga 0.4$ (Figure \ref{Interpolating_functions}), consistent with the estimate that $\alpha = 5/12$ and should certainly lie in the range $0.4 - 0.5$ \citep{Hajdukovic_2020}. Starting from the same initial position and velocity, this extra acceleration would cause the position of Saturn to deviate from conventional expectations by $7000 \left( \alpha/0.4 \right)$ km over 15 years (Section \ref{Epicyclic_approximation}). The Cassini mission at Saturn \citep{Matson_1992} has been tracked over roughly this timespan to a precision of 32 m \citep{Viswanathan_2017}. No significant deviations were found from its conventionally calculated trajectory, casting serious doubt on the GD model \citep{Iorio_2019}.

One possible complication is that in addition to the Solar dipole halo responsible for an extra Sunwards acceleration, each planet would have its own dipole halo \citep{Hajdukovic_2019}. However, we showed that these haloes are truncated when the Solar gravity dominates over that of the planet (Section \ref{Planetary_perturbations}). Consequently, the Solar dipole halo interior to the orbit of Saturn contains vastly more effective mass than the dipole halo of Jupiter, or indeed any other planet. Even if the effective masses of other planets were arbitrarily adjusted or new ones invented, their time-varying direction as viewed from Saturn makes it impossible for them to precisely cancel out an extra Sunwards acceleration. While this may be possible for a time-averaged quantity like the rate of Saturn's perihelion advance, there would still be km-scale deviations from its conventional trajectory during the course of an individual orbit. Since Cassini tracking data provides several thousand range measurements over half its orbital period, such deviations should have been detected. Moreover, the Solar dipole halo interior to Saturn has an effective mass of $0.25 M_\bigoplus$, vastly exceeding observational uncertainties on the masses of Solar System bodies and the effective masses of their dipole haloes.

The Galactic gravity causes the Solar dipole halo to depart from spherical symmetry at distances ${\ga 5}$~kAU. However, $\rho_{qv}$ in this region should have only a small tidal effect on the Solar System (Section \ref{Galactic_gravity}). Additionally, the force would not always be aligned with the Sun-Saturn line while Saturn completes half an orbit.

An extra Sunwards acceleration of order $a_{_0}$ is thus a strong prediction of the GD model for any planet in the Solar System. This prediction is robust to the possible presence of a ninth planet in the outer Solar System (Section \ref{Planet_9}). Since the anomalous acceleration should be the same for all planets, it cannot be consistently cancelled by tides raised by sources outside the Solar System.

Our work confirms the finding of \citet{Iorio_2019} that the GD model is falsified in its present form because it predicts unobserved effects within the Solar System that exceed observational upper limits by $\approx 5$ orders of magnitude. This falsification applies to any MOND-like force law with interpolating function given by Equation \ref{Interpolating_function_Hadjukovic} $-$ galactic rotation curve fits require $\alpha > 0.4$ while Solar System constraints imply $\alpha < 10^{-4}$ at very high significance. More generally, Solar System ephemerides rule out any theory that predicts an enhancement to gravity by $\ga 10^{-5} a_{_0}$ arbitrarily deep into the Newtonian regime. Even though MOND need not imply such an effect ($\nu$ can be 1 for high $g_{_N}$), \citet{Hees_2016} constrained its interpolating function based on the fact that the Galactic external field causes the Solar phantom dark matter halo to become non-spherical at large distances, imposing a tidal stress \citep{Blanchet_2011}.

We conclude that the gravitational dipole model proposed by \citet{Hajdukovic_2020} is unable to satisfy both galaxy rotation curve and Solar System constraints. The impressive successes of Milgromian dynamics must then be explained in some other way and other solutions must be found to the cosmological constant problem.

\section*{Acknowledgements}

IB is supported by an Alexander von Humboldt postdoctoral research fellowship. The authors are grateful to D. S. Hajdukovic for visiting them to discuss his GD model. They also thank the referee for comments which helped to improve this publication. The graphs were produced using \textsc{matlab}$^\text{\textregistered}$.

\bibliographystyle{mnras}
\bibliography{GDP_bbl}
\bsp
\label{lastpage}
\end{document}